\documentclass[aps,prd,showpacs,preprintnumbers,superscriptaddress,eqsecnum,nofootinbib]{revtex4}
\usepackage[dvips]{graphicx}
\usepackage{bm,latexsym,amsmath,amssymb,amsfonts}
\newcommand*{\cR}{{\cal R}}
\newcommand*{\cF}{{\cal F}}

\newcommand*{\cG}{{\cal G}}
\newcommand*{\cX}{{\cal X}}
\newcommand*{\cV}{{\cal V}}
\newcommand*{\D}{{\rm D}}
\newcommand*{\bb}{{\rm b}}
\begin{document}

\title{Brane cosmological solutions in six-dimensional warped flux compactifications}

\author{Tsutomu Kobayashi}
\email[Email: ]{tsutomu"at"gravity.phys.waseda.ac.jp}
\affiliation{Department of Physics, Waseda University, Okubo 3-4-1, Shinjuku, Tokyo 169-8555, Japan}

\author{Masato~Minamitsuji}
\email[Email: ]{masato"at"theorie.physik.uni-muenchen.de}
\affiliation{Arnold-Sommerfeld-Center for Theoretical Physics, Department f\"{u}r Physik, Ludwig-Maximilians-Universit\"{a}t, Theresienstr. 37, D-80333, Munich, Germany}

\begin{abstract}
We study cosmology on a conical brane in the six-dimensional
Einstein-Maxwell-dilaton system,
where the extra dimensions are compactified by a magnetic flux.
We systematically construct exact cosmological solutions
using the fact that
the system is equivalently described by $(6+n)$-dimensional pure
Einstein-Maxwell theory via dimensional reduction.
In particular, we find a power-law inflationary solution for a general dilatonic coupling.
When the dilatonic coupling is given by that of Nishino-Sezgin
chiral supergravity, this reduces to the known solution which is not inflating.
The power-law solution is shown to be
the late-time attractor.
We also investigate
cosmological tensor perturbations in this model
using the $(6+n)$-dimensional description.
We obtain the separable equation of motion and
find that there always exist a zero mode, while tachyonic modes are absent in the spectrum.
The mass spectrum of Kaluza-Klein modes is obtained numerically.
\end{abstract}

\pacs{04.50.+h, 98.80.Cq}
\preprint{WU-AP/266/07}
\preprint{LMU-ASC 33/07}
\maketitle

\section{Introduction}

Higher dimensional models have a longstanding history,
but there has in particular been a growing interest
since the discovery of D-branes in string theory,
leading to the concept of ``braneworld'' scenarios.
The braneworld framework
allows us to consider large (or even infinite)
extra dimensions due to localization of matter fields on the brane
\cite{Rubakov}.
The brane models with two extra-dimensions have drawn
considerable attention among others.
This is because
six-dimensional (6D) models may be able to eliminate the hierarchy problem
in particle phenomenology,
and at the same time bring us the possibility
to detect signatures of extra dimensions,
as the compactification radius can be of order 0.1\;mm~\cite{ADD}.
Moreover,
recent realizations of such brane models with
rugby-ball (or football) shaped extra-dimensions
provide a possible self-tuning mechanism
to resolve the cosmological constant problem~\cite{cc},
though the mechanism has been criticized for several reasons~\cite{cc2}.

In contrast to the 5D (codimension 1) brane models (e.g.,~\cite{RS,RScos}),
cosmological aspects of 6D models
have not been explored much (e.g.,~\cite{co2, papa}).
Taking into account the gravitational backreaction of branes,
one immediately faces the problem of the localization of matter:
it is difficult for a codimension 2 defect to accommodate
energy-momentum tensor different from pure tension.
This fact hampers attempts to construct 6D models for which
3-branes have Friedmann-Robertson-Walker geometry.
One will notice, however, that not only brane localized matter fields but also
those in the bulk can support a cosmic expansion of branes.
We focus here on this latter possibility.\footnote{One of the possible ways out is
to regularize conical deficits~\cite{Peloso1,Peloso2, PPZ, K-M, kalo, yama, bht}
so as to put ordinary matter on the brane.
Changing the bulk gravity theory (e.g.,~\cite{cz}) also helps.}

The purpose of this
paper is to present exact cosmological solutions 
in the presence of a scalar field, flux, and conical 3-branes
in six dimensions.
We will be considering the Einstein-Maxwell-dilaton system,
and for a certain choice of a parameter
our Lagrangian reduces to that of Nishino-Sezgin
chiral supergravity~\cite{N-S, S-S}.
In Ref.~\cite{Mukohyama}, Mukohyama et al. presented
a warped braneworld configuration
in a more simplified setup in Einstein-Maxwell theory.
The perturbation dynamics and stability issue have been extensively
studied in~\cite{Yoshiguchi, Sendouda, ksm}.
In the context of Nishino-Sezgin supergravity,
similar warped compactification solutions with Minkowski branes
were found in~\cite{Gibbons, A_et_al, Burgess1}
and de Sitter braneworlds were explored in \cite{dS}. 
While none of these solutions have a nontirivial time-dependence
of metric functions, Tolley et al.~\cite{Scaling} found cosmological scaling solutions in
6D chiral supergravity without showing
explicitly the metric of the internal space (see also Ref. \cite{Anchordoqui:2007sb}
for a numerical construction of cosmological solutions
in Nishino-Sezgin supergravity).
The solutions given in this paper are exact both
for the noncompact four dimensions and for the compact internal space.
After constructing our cosmological solutions,
we shall also discuss the behavior of tensor perturbations in the present model.


We will be taking the approach developed in Ref.~\cite{kt},
which was originally used to study
5D warped brane models with a scalar field
~\cite{5d1,5d2,5d3,5d4,koyamatakahashi,M-C}.
In~\cite{kt},
instead of solving the relevant 5D field equations,
the authors employed a $(5+n)$-dimensional equivalent description
{\em without} a scalar field.
This approach, with the extra $n$-dimensional space
playing effectively the role of the scalar field after
dimensional reduction,
greatly simplifies the analysis of cosmological perturbations
as well as constructing dynamical background solutions.
Here we
use a $(6+n)$-dimensional Einstein-Maxwell system
instead of working directly in the 6D model with a scalar field,
and show that
the same mathematical technique as~\cite{kt} successfully works.

The structure of the paper is as follows.
In the next section we introduce our dimensional reduction approach
and demonstrate how one can obtain a Minkowski braneworld
using the $(6+n)$-dimensional description.
In Sec.~\ref{background} we construct cosmological solutions
in the 6D braneworld.
The behavior of tensor perturbations
is discussed in Sec.~\ref{perturbations}.
Finally, our conclusions are summarized in Sec.~\ref{final}.


\section{Dimensional reduction approach to 6D braneworld models}
\label{reduction}

We are interested in the 6D system described by the action 
\begin{eqnarray}
S^{(6)}=\int d^6x\sqrt{-g}
\left[\frac{M_{(6)}^4}{2}\left(R[g]
-\partial_a\varphi\partial^a\varphi-2e^{-\gamma\varphi}\Lambda\right)
-\frac{1}{4}e^{\gamma\varphi}F_{ab}F^{ab}
\right], \label{6d_action}
\end{eqnarray}
with two or fewer branes.
Here $F_{ab}$ denotes the $U(1)$ field strength.
For $\gamma=1$ the action~(\ref{6d_action}) coincides with
the bosonic part of Nishino-Sezgin 
supergravity~\cite{N-S, S-S}
with some fields set to be zero consistently
(and with the identification of $\Lambda\to 2g^2_1M^4_{(6)}$, where
$g_1$ is the $U(1)$ gauge coupling).
In this paper we consider
more general cases and assume that the parameter $\gamma$ is arbitrary
in the range $0\leq\gamma\leq 1$.

The above Einstein-Maxwell-dilaton system with branes
has an equivalent $(6+n)$-dimensional description
in terms of pure Einstein-Maxwell theory.
To show this, we start with the $(6+n)$-dimensional Einstein-Maxwell action
\begin{eqnarray}
S^{(6+n)}=\int d^{6+n}\cX\sqrt{-\cG}
\left[\frac{M^{4+n}_{(6+n)}}{2}\left(
\cR\left[\cG\right]-2\Lambda\right)-\frac{1}{4}\cF_{MN}\cF^{MN}\right],
\label{6+n-d-action}
\end{eqnarray}
and the $(4+n)$-dimensional pure tension branes $(i=1, 2)$
\begin{eqnarray}
S^{(4+n)}_{\bb i}=-\int d^{4+n}{\cal X}\sqrt{-\cG_{\text{b}i}}\,{\cal T}_i,
\label{(4+n)brane_action}
\end{eqnarray}
where ${\cal G}_{\text{b}i}$ is the determinant of the induced metric
on the brane.

Let us consider the metric of the form
\begin{eqnarray}
\cG_{MN}d\cX^Md\cX^N=
\underbrace{e^{-n\phi(x)/2}g_{ab}(x)dx^adx^b}_{6\D}
+\underbrace{e^{2\phi(x)}\delta_{mn}dy^mdy^n}_{n\D},
\label{metric_form_6+n}
\end{eqnarray}
and the field strength with
\begin{eqnarray}
\cF_{ab}=\cF_{ab}(x)\quad\text{and}\quad\cF_{mM}=0.\label{fstr}
\end{eqnarray}
With the above ansatz, dimensional reduction yields
\begin{eqnarray}
S^{(6)}=\int d^6x\sqrt{-g}\left[ \frac{M^4_{(6)}}{2}\left(
R[g]-\frac{n(n+4)}{4}\partial_a\phi\partial^a\phi-2e^{-n\phi/2}\Lambda\right)
-\frac{1}{4}e^{n\phi/2}F_{ab}F^{ab}
\right],\label{reduction_to_6d_action}
\end{eqnarray}
where $M_{(6)}^4:=M_{(6+n)}^{4+n}\cV_{(n)}$, $F_{ab}:=\cF_{ab}\cV_{(n)}^{1/2}$,
and $\cV_{(n)}$ is the volume of the $n$-dimensional flat space.
Note that in~(\ref{reduction_to_6d_action}) indices are raised and lowered with $g_{ab}$.
Now, defining
\begin{eqnarray}
\varphi:=\frac{\sqrt{n(n+4)}}{2}\,\phi,
\quad
\gamma:=\sqrt{\frac{n}{4+n}},
\end{eqnarray}
we see that (\ref{reduction_to_6d_action}) is identical to the
action~(\ref{6d_action}).
We also define the 3-brane tension as $T_i:={\cal T}_i \cV_{(n)}$. Then
the brane action~(\ref{(4+n)brane_action}) reduces to
\begin{eqnarray}
S_{\bb i}^{(4)}= - \int d^4x \sqrt{-q_i}\,T_i,\label{4dpuret}
\end{eqnarray}
where $q_i$ is the determinant of the induced metric on the 3-brane.
The tension of the 3-brane
does not couple to the dilatonic scalar field after dimensional reduction.
Thus, we may use the $(6+n)$-dimensional
pure Einstein-Maxwell description~(\ref{reduction_to_6d_action})
with~(\ref{(4+n)brane_action}) instead of working directly
in the 6D Einstein-Maxwell-dilaton system~(\ref{6d_action})
with tensional branes~(\ref{4dpuret}).
Given a metric $\cG_{MN}$ of the form~(\ref{metric_form_6+n})
and a gauge field that satisfies~(\ref{fstr}),
one automatically obtains $g_{ab}$, $F_{ab}$, and $\varphi$ that solve the action~(\ref{6d_action}).
Since the action~(\ref{6+n-d-action}) is more simple than~(\ref{6d_action}),
it is often easier to find a solution of~(\ref{6+n-d-action}).
Note that in the action~(\ref{reduction_to_6d_action})
$n$ is an arbitrary parameter and may no longer be an integer.
The right signature for the scalar kinetic term [$n(n+4)\geq 0$]
ensures that $\gamma^2$ is nonnegative.

The field equations derived directly from the action~(\ref{6d_action})
possess the scaling symmetry
\begin{eqnarray}
g_{ab}\to u \,g_{ab}, \qquad \varphi\to\varphi+\gamma^{-1}\ln u,
\end{eqnarray}
with constant $u$.
Note that this is not a symmetry of the action, as $S^{(6)}\to u^2S^{(6)}$.
From the $(6+n)$-dimensional point of view, this symmetry is manifest
because we may write
\begin{eqnarray}
\cG_{MN}d\cX^Md\cX^N=e^{-n\phi/2-\ln u}ug_{ab}dx^adx^b+\cdots,
\end{eqnarray}
and use $ug_{ab}$ and $\phi+(2/n)\ln u$ instead of $g_{ab}$ and $\phi$.

We would like to obtain six-dimensional
axisymmetric warped compactification solutions with conical branes
at the points where the geometry pinches off. To this end
we make use of the $(6+n)$-dimensional generalization of
the exact solution given in Ref.~\cite{Mukohyama},
which is obtained by a double Wick rotation
from the higher dimensional Reissner-Nordstr\"{o}m solution.
The metric is written as
\begin{eqnarray}
\cG_{MN}d\cX^Md\cX^N=\underbrace{r^2 \bar g_{\alpha\beta}dz^{\alpha}dz^{\beta}}_{(4+n)\D}
+\frac{1}{2\Lambda}\left[
\frac{dr^2}{\tilde f(r)}+\tilde\beta^2\tilde f(r)d\theta^2\right],
\end{eqnarray}
where
\begin{eqnarray}
\tilde f(r)=\frac{\lambda}{(2+n)(3+n)\Lambda}+\frac{1}{4+n}\left[
\frac{C}{r^{3+n}}-\frac{r^2}{5+n}-\frac{Q^2}{2(3+n)M^{4+n}_{(6+n)}
\cV_{(n)}\Lambda}\frac{1}{r^{6+2n}}
\right],
\end{eqnarray}
and $\bar g_{\alpha\beta}$
is {\em any} metric that solves the $(4+n)$-dimensional field equations
\begin{eqnarray}
\bar R_{\alpha\beta}[\bar g]=\frac{2}{2+n}\lambda \bar g_{\alpha\beta}.
\label{4+n_field_eq}
\end{eqnarray}
The only nonvanishing component of the field strength is given by
\begin{eqnarray}
\cF_{r\theta}=\frac{Q\tilde\beta}{2\Lambda\sqrt{\cV_{(n)}}}\frac{1}{r^{4+n}}.
\end{eqnarray}
The constant $\tilde\beta$
can be absorbed in the angular coordinate $\theta$, but we keep it in the metric.

We assume that $\tilde f(r)$ has two positive roots,
$r_+$ and $r_-=\alpha r_+$ with $0<\alpha\leq 1$.
In terms of the new coordinate defined by $\chi:=r/r_+$,
the metric function can be written as $\tilde f =r_+^2 f(\chi)$, where
\begin{eqnarray}
f(\chi)&=&\frac{1}{(4+n)(5+n)}\left[
-\chi^2+
\frac{1-\alpha^{8+2n}}{1-\alpha^{3+n}}\frac{1}{\chi^{3+n}}
-\frac{\alpha^{3+n}(1-\alpha^{5+n})}{1-\alpha^{3+n}}\frac{1}{\chi^{6+2n}}
\right]
\nonumber\\&&\qquad\qquad\qquad
+\frac{\lambda}{(2+n)(3+n)\Lambda r^2_+}\left[1-\frac{1}{\chi^{3+n}}\right]
\left[1-\frac{\alpha^{3+n}}{\chi^{3+n}}\right].
\label{new_f}
\end{eqnarray}
The parameter $\alpha$ characterizes warping of the bulk.
In the case of $\alpha=1$, the bulk is given by a two-dimensional sphere
and is often referred to as the rugby-ball compactification.

The rescaling $r_+^2\bar g_{\alpha\beta}dz^{\alpha}dz^{\beta}\to \bar g_{\alpha\beta}dz^{\alpha}dz^{\beta}$
and $\lambda/r_+^2\to \lambda$ allows us to write
\begin{eqnarray}
\cG_{MN}d\cX^Md\cX^N=\chi^2\bar g_{\alpha\beta}dz^{\alpha}dz^{\beta}
+\frac{1}{2\Lambda}\left[
\frac{d\chi^2}{f(\chi)}+\beta^2f(\chi)d\theta^2\right],\label{intermsofchi}
\end{eqnarray}
and
\begin{eqnarray}
\cF_{\chi\theta}=\frac{ M^{2+n/2}_{(6+n)}\beta}{\sqrt{2\Lambda}}\frac{\hat Q}{\chi^{4+n}},
\label{F-chi-th}
\end{eqnarray}
where
\begin{eqnarray}
\hat Q:=\left[
\frac{3+n}{5+n}\frac{1-\alpha^{5+n}}{1-\alpha^{3+n}}
-\frac{4+n}{2+n}\frac{\lambda}{\Lambda}
\right]^{1/2}\alpha^{(3+n)/2}. \label{def_Q}
\end{eqnarray}
With this rescaling, $r_+$ in Eq.~(\ref{new_f}) is absorbed in $\lambda$
and therefore the solution does not have an explicit dependence on $r_+$,
while $\bar g_{\alpha\beta}$ still solves Eq.~(\ref{4+n_field_eq}).

It is clear from
Eq.~(\ref{def_Q}) that $\lambda$ is bounded from above:
\begin{eqnarray}
\lambda\leq \lambda_{\text{max}}(\alpha):=
\frac{(2+n)(3+n)}{(4+n)(5+n)}\frac{1-\alpha^{5+n}}{1-\alpha^{3+n}}\Lambda.
\label{upperboundlambda}
\end{eqnarray}
A large value of $\lambda$ reduces the strength of flux,
and the flux vanishes if the bound is saturated.
Such a bound on the brane curvature scale
can also be observed
in 6D flux compactification models with de Sitter branes~\cite{Peloso2, ksm}.

After dimensional reduction $n$ is a parameter of the 6D theory
which is directly related to the dilatonic coupling. In particular,
Nishino-Sezgin supergravity ($\gamma=1$) is reproduced by taking $n\to\infty$.
In this limit the metric function $f(\chi)$ is apparently singular.
This is due to a bad choice of the coordinate in the extra direction.
A regular expression is obtained by using $\xi:=\chi^{1+n/2}$ and $\hat\alpha:=\alpha^{1+n/2}$,
in terms of which one can write the metric in $(6+n)$ dimensions as
\begin{eqnarray}
\cG_{MN}d\cX^Md\cX^N=\xi^{4/(2+n)}\bar g_{\alpha\beta}dz^{\alpha}dz^{\beta}
+\frac{\xi^{-n/(2+n)}}{2\Lambda}\left[
\frac{d\xi^2}{h(\xi)}+\hat\beta^2h(\xi)d\theta^2\right],
\label{intermsofxi}
\end{eqnarray}
where
\begin{eqnarray}
h(\xi)&:=&\frac{1}{2(1+\omega)(2+3\omega)}\left[
-\xi^{1+\omega}
+\frac{1-\hat\alpha^{4(1+\omega)}}{1-\hat\alpha^{2+\omega}}
\frac{1}{\xi^{1+2\omega}}-
\frac{\hat\alpha^{2+\omega}(1-\hat\alpha^{2+3\omega})}{1-\hat\alpha^{2+\omega}}
\frac{1}{\xi^{3(1+\omega)}}
\right]
\nonumber\\&&\qquad\qquad
+\frac{\lambda}{\Lambda}\frac{\xi^{1-\omega}}{2(2+\omega)}
\left[ 1-\frac{1}{\xi^{2+\omega}} \right]
\left[ 1-\frac{\hat\alpha^{2+\omega}}{\xi^{2+\omega}} \right],
\end{eqnarray}
and $\hat\beta:=\omega\beta$ with $\omega:=(1+n/2)^{-1}$.
Clearly, $h$ is regular for $n\to\infty$ ($\omega=0$).
The field strength is now given by
\begin{eqnarray}
\cF_{\xi\theta}=\frac{ M^{2+n/2}_{(6+n)}\hat\beta}{\sqrt{2\Lambda}}
\frac{\hat Q}{\xi^{(8+3n)/(2+n)}}.\label{fs}
\end{eqnarray}

We assume that $\theta$ has period $2\pi$.
The constant $\beta$ (or $\hat\beta$) controls deficit angles
at $\chi=1$ ($\xi=1$) and $\chi=\alpha$ ($\xi=\hat\alpha$), which are given, respectively, by
\begin{eqnarray}
\delta_1=2\pi\left[1-\frac{\beta}{2}\left|f'(1)\right|\right]
=2\pi\left[1-\frac{\hat\beta}{2}
\left|h'(1)\right|\right],
\quad
\delta_2
=2\pi\left[1-\frac{\beta}{2}\left|f'(\alpha)\right|\right]=2\pi\left[1-\frac{\hat\beta}{2}
\left|h'(\hat\alpha)\right|\right].
\end{eqnarray}
Here the prime stands for a derivative with respect to the argument.
The conical deficit corresponds to a codimension two brane and the tension is given by
${\cal T}_i/M^{4+n}_{(6+n)} = \delta_i$.
Whereas in general there are two branes in the present model,
one can choose $\beta$ so that one of the deficit angles vanishes. 
However, for the warped $\alpha\neq1$ case
it is not possible to make both of the deficit angles zero.
Hence, the warped model includes at least one brane.
The deficit angles satisfy the constraint
\begin{eqnarray}
\frac{2\pi-\delta_2}{2\pi-\delta_1} = \left|\frac{h'(\hat\alpha)}{h'(1)}\right|
\;
\left[
=\left|\frac{f'( \alpha)}{f'(1)}\right|
\right],
\end{eqnarray}
which therefore places the constraint between the brane tensions.

It is instructive to present here the simplest example: a Minkowski braneworld.
The desired solution is generated from
the $(4+n)$-dimensional Minkowski metric:
$\bar g_{\alpha\beta}dz^{\alpha}dz^{\beta}=\eta_{\mu\nu}dx^{\mu}dx^{\nu}+\delta_{mn}dy^mdy^n$
(and hence $\lambda=0$).
We identify $e^{\phi}=\chi=\xi^{2/(2+n)}$ and then
the 6D metric is
\begin{eqnarray}
g_{ab}dx^adx^b=\xi^{2/(1+\gamma^2)}\eta_{\mu\nu}dx^{\mu}dx^{\nu}
+\frac{1}{2\Lambda}\left[\frac{d\xi^2}{h(\xi)}+\hat\beta^2
h(\xi)d\theta^2\right].
\end{eqnarray}
The field strength and the scalar field is given, respectively, by
\begin{eqnarray}
F_{\xi\theta}&=&\frac{M_{(6)}^2\hat\beta}{\sqrt{2\Lambda}}
\frac{\hat Q}{\xi^{2(2+\gamma^2)/(1+\gamma^2)}},
\label{fsMink}
\\
\varphi &= &\frac{2\gamma}{1+\gamma^2}\ln\xi.
\end{eqnarray}
Since ${\cal T}_i/M^{4+n}_{(6+n)}=T_i/M^{4}_{(6)}$, the conical deficit and the 3-brane tension
are related by $T_i/ M^{4}_{(6)}=\delta_i$.
One can check that the above configuration reduces
to the known solution in the Einstein-Maxwell model~\cite{Yoshiguchi, Sendouda}
and in Nishino-Sezgin supergravity~\cite{Gibbons, A_et_al, Burgess1}
for the corresponding value of $\gamma$.


\section{Exact cosmological solutions}\label{background}

Cosmological solutions can be systematically generated from the following Kasner-type
metric,\footnote{We derive the main results in this section
assuming $n>0$ and hence $\gamma\neq 0$.
Obviously, the $n=0$ case does not admit the nontrivial Kasner-type solution
and only the de Sitter geometry (as in~\cite{Mukohyama}) is allowed on the brane.
}
\begin{eqnarray}
\bar g_{\alpha\beta} dz^{\alpha}dz^{\beta}=-dt^2+
\underbrace{e^{2A(t)}\delta_{ij}dx^idx^j}_{\text{3D}}
+
\underbrace{e^{2B(t)}\delta_{mn}dx^mdx^n}_{n\text{D}}.
\label{4+n_met}
\end{eqnarray}
The explicit form of the time-dependent metric functions $A(t)$ and $B(t)$
is derived in detail in Appendix~\ref{App:Kasner}.
Looking at
Eq.~(\ref{metric_form_6+n}) and the metric~(\ref{intermsofchi}) [or~(\ref{intermsofxi})]
with~(\ref{4+n_met}),
we identify $e^{\phi}=\chi e^{B}=\xi^{2/(2+n)}e^{B}$.
Then the 6D metric is found to be
\begin{eqnarray}
g_{ab}dx^adx^b
&=&\xi^{2/(1+\gamma^2)}
\left[-e^{nB/2}dt^2+e^{2A+nB/2}\delta_{ij}dx^idx^j\right]
+\frac{e^{nB/2}}{2\Lambda}
\left[\frac{d\xi^2}{h(\xi)}+\hat \beta^2 h(\xi) d\theta^2\right]
\nonumber\\
&=&
\xi^{2/(1+\gamma^2)}
\left[-d\tau^2+a^2(\tau)\delta_{ij}dx^idx^j\right]
+\frac{b^{2}(\tau)}{2\Lambda}\left[\frac{d\xi^2}{h(\xi)}
+\hat\beta^2h(\xi)d\theta^2\right],
\label{new_6d_sol}
\end{eqnarray}
where we have introduced the proper time on the brane, $\tau=\int e^{nB/4}dt$,
and defined the scale factors of the external and internal spaces as
$a(\tau):=e^{A+nB/4}$ and $b(\tau):=e^{nB/4}$, respectively.
The dilaton is given by
\begin{eqnarray}
\varphi(\tau, \xi)=\frac{2}{\gamma}\ln b(\tau)
+\frac{2 \gamma}{1+\gamma^2}\ln\xi.
\end{eqnarray}
The field strength $F_{\xi\theta}$ is the same
as that in the static solution~(\ref{fsMink}).

Let us first consider the simple case where the metric functions are
given by
\begin{eqnarray}
A(t)=H_0 t+A_0, \quad B(t)=H_0t+B_0, \label{sol_dS}
\end{eqnarray}
with $H_0^2:=2\lambda/[(3+n)(2+n)]\neq0$.
In this case it is easy to see
that
\begin{eqnarray}
a(\tau)\propto \tau^{1/\gamma^2}, \quad b(\tau)\propto\tau,
\label{power-law}
\end{eqnarray}
leading to power-law inflation for $\gamma<1$.
However, the Nishino-Sezgin supergravity model, $\gamma=1$, 
does not give an accelerated expansion.
The $\gamma=1$ solution was first found by Tolley et al.~\cite{Scaling}.
While they have not presented an explicit form of the metric
of the two-dimensional internal space,
Eq.~(\ref{new_6d_sol}) shows the full spacetime metric in six dimensions.

In addition to the above trivial solution, we have another type of solutions, 
\begin{eqnarray}
e^{3A+nB}&=&C_3\sinh\left[\sqrt{\frac{3+n}{2+n}}\sqrt{2\lambda}\,(t-t_0)\right],
\label{nontr1}\\
e^{A-B}&=&C_4\left\{\tanh\left[\sqrt{\frac{3+n}{2+n}}\sqrt{\frac{\lambda}{2}}\,(t-t_0)\right]
\right\}^{\pm\sqrt{(2+n)/3n}}.\label{nontr2}
\end{eqnarray}
At early times, $t-t_0\sim0$, we have
$a\sim \tau^{p_{\pm}}$ and $b\sim\tau^{q_{\pm}}$ where
\begin{eqnarray}
p_{\pm}=
\frac{ 3(2+\gamma^2)\pm 2\sqrt{6}\gamma\sqrt{1+\gamma^2}}
      {3(6+5\gamma^2)},
\quad
q_{\pm}=
      \frac{\gamma^2\mp \sqrt{6} \gamma\sqrt{1+\gamma^2}}
           {6+5\gamma^2}.
\label{early}
\end{eqnarray}
The ranges of $p_{\pm}$ and $q_{\pm}$ are
$1/3<p_+\leq(9+4\sqrt{3})/33\approx0.48$,
$1/3>p_-\geq(9-4\sqrt{3})/33\approx0.063$,
$0>q_+\geq(1-2\sqrt{3})/11\approx -0.22$, and
$0<q_-\leq(1+2\sqrt{3})/11\approx 0.41$.
Note that $q_+$ is negative and hence in this case
the compact two-dimensional space is initially contracting.
At late times, $t\to\infty$, we have $a\sim\tau^{1/\gamma^2}$
and $b\sim\tau$. Therefore,
the solution~(\ref{power-law}) is the late-time attractor.
This attractor behavior on the 4D brane
emerges as a consequence of
Wald's cosmic no hair theorem~\cite{wald} for the $(4+n)$-dimensional metric $\bar g_{\alpha\beta}$:
the initially expanding Bianchi-type metric~(\ref{4+n_met})
isotropizes in the presence of the positive
cosmological constant $\lambda$.

The cosmological dynamics can also be understood
in terms of the 4D effective theory in the zero-mode sector.
In Appendix~\ref{App:effective4d} we show that the induced metric on the brane
can be derived from the Einstein-scalar field action with an exponential potential
up to conformal transformations.
Given generic initial conditions for the ``position'' and ``velocity'' of the scalar field,
the scale factor and the time-dependence of $\varphi$ are
obtained from~(\ref{nontr1}) and~(\ref{nontr2}).
In this case the two of the three integration constants correspond to
the initial position and velocity of the scalar field, whereas
the third integration constant corresponds to the overall normalization
of the scale factor.
The solution~(\ref{power-law}) is realized
for the special choice of the initial velocity (depending on the initial position),
as~(\ref{sol_dS}) has only two integration constants.
The power-law solution obtained from~(\ref{sol_dS}) is the late-time attractor
for the exponential potential of the scalar field.
The cosmic expansion looks very different in a different frame,
but the discussion here in terms of the 4D effective theory~(\ref{effective4d_conf})
is still helpful.

The dynamics of the $\lambda=0$ model is the same as the early-time
behavior of the general $\lambda\neq0$ case:
$a\sim\tau^{p_{\pm}}$ and $b\sim\tau^{q_{\pm}}$.
From Eq.~(\ref{early}) we see that
the $\gamma=1$ case gives the scale factor identical to
that in~\cite{MN1, MN2}.
One should note here that the geometry on the brane is independent of 
the warping parameter $\alpha$
and the brane tension.
They only affect the bulk geometry. 
For this reason, it is not surprising that
we can reproduce the scale factor of
the model without warping and branes~\cite{MN1, MN2} just by taking $\gamma=1$.


\section{Cosmological tensor perturbations}\label{perturbations}

We will be discussing the behavior of perturbations
in the cosmological background solution constructed in the previous section.
Here we make use of
the $(6+n)$-dimensional description again,
and focus on the axisymmetric tensor perturbations.\footnote{Perturbations
are decomposed into scalar, vector, and tensor
parts according to their transformation properties with respect to
the three-dimensional space. Scalar and vector perturbations
will be investigated in a forthcoming paper.}
The perturbed metric is given by
\begin{eqnarray}
\left(\cG_{MN}^{(0)}+\delta\cG_{MN}^{(1)}\right)d\cX^Md\cX^N
=\chi^2\left[-dt^2+e^{2A}\left(\delta_{ij}+h_{ij}\right)dx^idx^j+e^{2B}\delta_{mn}dy^mdy^n\right]
+\frac{1}{2\Lambda}\left[\frac{d\chi^2}{f}+\beta^2fd\theta^2\right],
\end{eqnarray}
where $h_{ij}=h_{ij}(t, \mathbf{x}, \chi)$ and $h_{i}^{~i}=\partial^jh_{ij}=0$.
The ($6+n$)-dimensional Einstein equations give
\begin{eqnarray}
e^{-3A-nB}\partial_t\left(e^{3A+nB}\partial_{t}h_{ij}\right)-e^{-2A}\nabla^2h_{ij}
=2\Lambda\chi^{-2-n}\partial_{\chi}\left(\chi^{4+n}f \partial_{\chi}h_{ij}\right),
\end{eqnarray}
which is separable.
In the Fourier space, the general solution can be written as
\begin{eqnarray}
h_{ij} =\sum_m \psi_{m}(t)\Omega_{m}(\chi) e^{\text{TT}}_{ij}(\mathbf{k}; \mathbf{x}),
\end{eqnarray}
where $e^{\text{TT}}_{ij}$ is the transverse traceless tensor harmonics which satisfies
$\nabla^2e_{ij}^{\text{TT}}=-k^2e_{ij}^{\text{TT}}$.

The mode solutions $\psi_m$ and $\Omega_m$ obey
\begin{eqnarray}
\left[\frac{d^2}{dt^2}+\left(3\dot A+n\dot B\right)\frac{d}{dt}
+\frac{k^2}{e^{2A}}+\mu^2_m\right]\psi_m&=&0,
\label{time-mode}
\\
\frac{1}{\chi^{2+n}}\frac{d}{d\chi}
\left(\chi^{4+n}f\frac{d}{d\chi}\Omega_m\right)+
\frac{\mu_m^2}{2\Lambda}\Omega_m&=&0,
\label{extra-mode}
\end{eqnarray}
where
the dot denotes a derivative with respect to $t$ and
$\mu_m^2$ is a separation eigenvalue.
At the poles the metric function $f$ vanishes.
Therefore, the boundary conditions for $\Omega_m$ are
\begin{eqnarray}
\chi^2 f'\Omega'_m+\frac{\mu^2_m}{2\Lambda}\Omega_m=0
\qquad\text{at}\qquad \chi=1, ~\alpha. \label{bc_Omega}
\end{eqnarray}
Two things should be remarked here.
First, there always exist a zero mode ($\mu_0^2=0$):
\begin{eqnarray}
\Omega_0=\text{const.}
\end{eqnarray}
Second,
there are no tachyonic modes ($\mu^2_m<0$) in the spectrum.
This can be shown as follows.
Using Eq.~(\ref{extra-mode}) we obtain
\begin{eqnarray}
\frac{\mu^2_m}{2\Lambda}\int^1_{\alpha}\chi^{2+n}\Omega^2_md\chi
&=&-\int^1_{\alpha}\Omega_m\left(\chi^{4+n}f \Omega_m'\right)'d\chi
=\int^1_{\alpha}\chi^{4+n} f\left(\Omega_m'\right)^2d\chi\geq0,
\end{eqnarray}
where we used $f(1)=f(\alpha)=0$ and assumed the regularity of $\Omega_m$ and $\Omega_m'$
at the boundaries. Thus, $\mu^2_m$ is not negative. The equality holds
only for $\Omega_m'=0$, which is definitely the zero mode.

Eq.~(\ref{time-mode}) can be written
in terms of the conformal time coordinate $\eta:=\int e^{-A}dt$ as
\begin{eqnarray}
\left[\frac{d^2}{d\eta^2}
+\left(2{\cal H}+\frac{n}{2} \frac{dB}{d\eta}\right)\frac{d}{d\eta}+
k^2+e^{-nB/2}\mu_m^2a^2\right]
\psi_m=0, \label{time-mode-con}
\end{eqnarray}
where ${\cal H}:=a^{-1}da/d\eta$ is the conformal Hubble parameter on the brane.
We see from Eq.~(\ref{time-mode-con}) that the Kaluza-Klein (KK) mass with respect
to observers on the 4D brane is time-dependent and is given by
\begin{eqnarray}
M_m^2(\eta):=e^{-nB(\eta)/2}\mu_m^2.
\end{eqnarray}
As for the zero mode, Eq.~(\ref{time-mode-con}) gives the
standard 4D equation for a massless graviton
except for the extra term $dB/d\eta$.
The presence of this term reflects the fact that the
effective theory for the zero-mode sector is given by a scalar-tensor gravity,
as is argued in~Appendix~\ref{App:effective4d}.

We are mainly interested
in the inflationary attractor solution~(\ref{sol_dS}). In this case,
Eq.~(\ref{time-mode-con}) reduces to
\begin{eqnarray}
\left(\frac{d^2}{d\eta^2}-\frac{2+n}{\eta}\frac{d}{d\eta}+k^2+\frac{\mu^2_m}{H_0^2\eta^2}
\right)\psi_m=0,
\end{eqnarray}
which can be solved exactly for any value of $\mu_m^2$.
The general solution is given by
\begin{eqnarray}
\psi_m &=& (-k\eta)^{(3+n)/2}Z_{i\nu_m}(-k\eta), \label{gen_bes}
\end{eqnarray}
where $Z_{\nu}$ is a linear combination of Bessel functions of order $\nu$ and
\begin{eqnarray}
\nu_m^2:=\frac{\mu_m^2}{H_0^2}-\frac{(3+n)^2}{4}.
\end{eqnarray}
The late-time asymptotic behavior of~(\ref{gen_bes}) depends on the value of $\mu^2_m$.
``Heavy'' modes with $\mu^2_m\geq\mu^2_c$, where
\begin{eqnarray}
\mu_c^2:=\frac{(3+n)\lambda}{2(2+n)},
\end{eqnarray}
decay rapidly as $|\psi_m|\sim a^{-2(3+n)/(4+n)}$ at late times ($-k\eta\to 0$).
``Light'' modes with $\mu^2_m<\mu^2_c$ decay more slowly,
and for $\mu^2_m\ll \mu^2_c$ we have $|\psi_m|\sim a^{-\mu_m^2/\mu_c^2\cdot(3+n)/(4+n)}$.
In particular, the zero-mode fluctuations are frozen, $\psi_0\sim$ const., at late times.
Since
\begin{eqnarray}
\frac{M_m^2a^2}{{\cal H}^2}=\left[\frac{2(3+n)}{(4+n)}\right]^2 \frac{\mu^2_m}{\mu^2_c},
\end{eqnarray}
heavy (light) modes in the $(6+n)$-dimensional description
are also heavy (light) for a 4D brane observer.

\begin{figure}[t]
  \begin{center}
    \includegraphics[keepaspectratio=true,height=50mm]{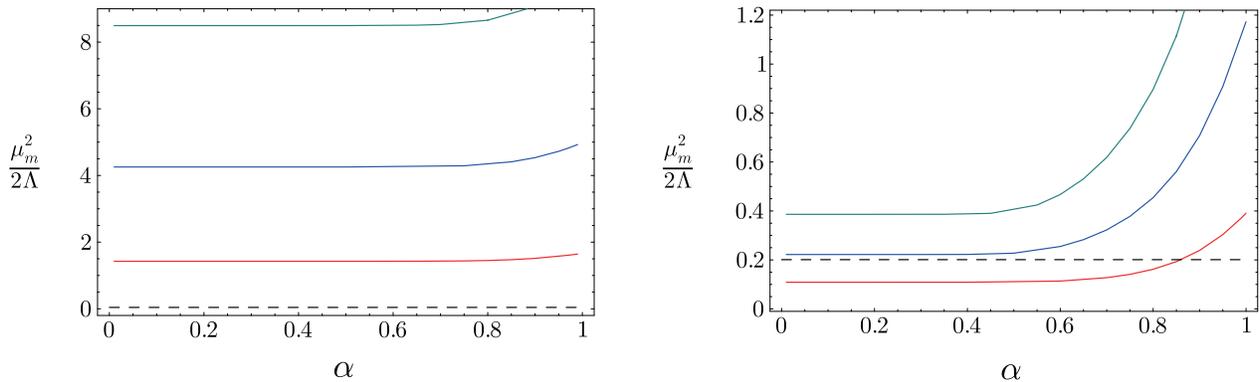}
  \end{center}
  \caption{Eigenvalues of the first three Kaluza-Klein modes, $\mu^2_1$, $\mu_2^2$, and $\mu^2_3$,
  versus $\alpha$ for $n=10$. The parameter is given by
  $\lambda/\Lambda=0.156$ (left) and $\lambda/\Lambda=0.74256$ (right).
  The dashed line indicates $\mu^2_c$.}
  \label{fig:kkmass123.eps}
\end{figure}

Now we are to determine
the mass spectrum of KK modes $\{\mu_1^2, \mu_2^2, \cdots\}$ numerically.
The spectrum
depends on $n$ and the parameters of the background solution, $\alpha$ and $\lambda$.
Given a set of parameters, we numerically solve Eq.~(\ref{extra-mode})
supplemented
with the condition $\Omega(1)=1$ and
$\Omega'(1)=-\mu^2/[2\Lambda f'(1)]$.
Then we have
\begin{eqnarray}
I(\mu^2):=\left[\alpha^2f'\Omega'+\frac{\mu^2}{2\Lambda}\Omega\right]_{\chi=\alpha},
\end{eqnarray}
as a function of $\mu^2$. The mass spectrum is given by zeros of $I(\mu^2)$.

\begin{figure}[t]
  \begin{center}
    \includegraphics[keepaspectratio=true,height=50mm]{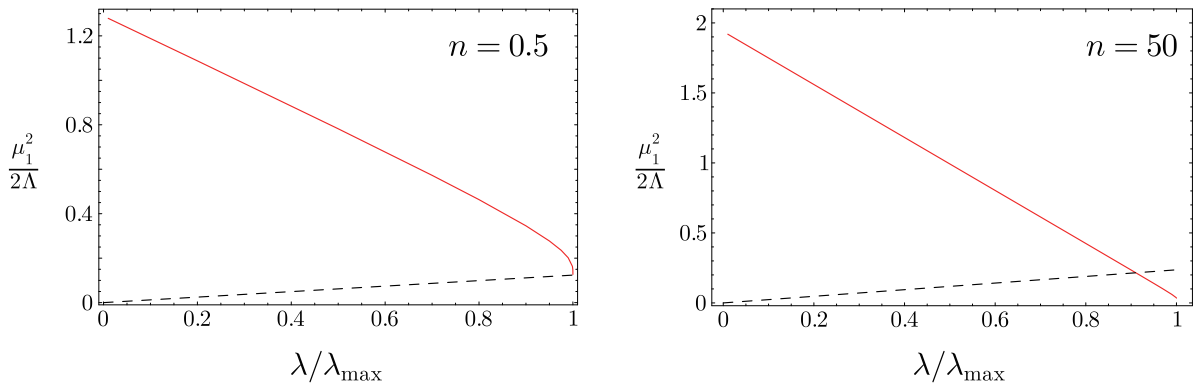}
  \end{center}
  \caption{First Kaluza-Klien eigenvalue in the limit of $\alpha\to0$ versus
  $\lambda/\lambda_{\text{max}}(0)$.
  The dashed line indicates $\mu_c^2$.}
  \label{fig:L-kk.eps}
\end{figure}

Examples of $\mu^2_m$ are shown in Fig.~\ref{fig:kkmass123.eps}.
As is clear, $\mu_m^2$ is an increasing function of $\alpha$.
One finds that $\mu_1^2$ can be smaller than $\mu_c^2$
for a relatively large value of $\lambda$.
To see this more closely, we plot in Fig.~\ref{fig:L-kk.eps}
the lightest KK mass $\mu^2_1$ in the limit of $\alpha\to 0$ as a function of $\lambda$.
We find that for $n\sim 1$ we have $\mu_1^2\sim \mu_c^2$ when
$\lambda$ is close to its maximum value, while
larger $n$ tends to give a very light mode for large $\lambda$.

If we have very light modes in the spectrum,
such modes can be excited during the period of inflation and decay slowly.
This is certainly an interesting situation, but is not favored for the following reason.
According to Ref.~\cite{dS:unstable},
a large expansion rate in the external coordinates generally
has the effect of destablizing
de Sitter compactifications.
In fact, the presence of $\lambda$
(corresponding to the $(4+n)$-dimensional Hubble rate)
decreases the strength of flux as seen in Eq.~(\ref{def_Q}).
It was also confirmed directly in~\cite{ksm} that
in the 6D flux compactification model
there appears a tachyonic mode in the scalar sector of perturbations
when the expansion on the brane is too large (but still less than the 
upper bound given by Eq. (\ref{upperboundlambda})
with $n=0$).\footnote{We should remark that in~\cite{ksm} scalar,
vector, and tensor perturbations are defined with respect to the 4D de Sitter spacetime}
From this we may expect that the $(6+n)$-dimensional Einstein-Maxwell model
will also be unstable for large $\lambda$, leading to
the instability of the 6D Einstein-Maxwell-dilaton model
since the two descriptions are equivalent.
In light of this, it is natural to consider that
smaller $\lambda$ is favored also in the present model.
For sufficiently small $\lambda$ all the KK modes are heavy
enough to decay rapidly at late times.

For understanding the behavior of KK perturbations in more detail,
it would be helpful to provide an analytic argument on the mode functions
and eigenvalues in the special case of $\alpha=1$.
Rewriting Eq.~(\ref{extra-mode}) in terms
of the new coordinate $w:=(2\chi-1-\alpha)/(1-\alpha)$
and then taking the limit $\alpha\to 1$, we get the equation
\begin{eqnarray}
\left(1-\frac{3+n}{4+n}\Delta\right)\frac{d}{d w}
\left[(1-w^2)\frac{d}{dw}\Omega_m\right]+\frac{\mu^2_m}{2\Lambda}\Omega_m=0,
\end{eqnarray}
subject to the boundary conditions
\begin{eqnarray}
\mp 2\left(1-\frac{3+n}{4+n}\Delta\right)
\frac{d\Omega_m}{dw}+\frac{\mu^2_m}{2\Lambda}\Omega_m=0
\quad \text{at} \quad w=\pm1,
\end{eqnarray}
where $\Delta:=\lambda/\lambda_{\text{max}}(1)$.
The mode solution is given up to the overall normalization by
\begin{eqnarray}
\Omega_m = P_m(w),
\quad \frac{\mu^2_m}{2\Lambda}=m(m+1)\left(
1-\frac{3+n}{4+n}\Delta\right),
\label{a0mass}
\end{eqnarray}
where $P_m$ is the Legendre function of the first kind of order $m$ and $m=0, 1, 2, \cdots$.
We checked that this analytic result agrees with our numerical calculations
in the $\alpha\to1$ limit.

It follows from the mass spectrum~(\ref{a0mass}) that
\begin{eqnarray}
\frac{\mu^2_m}{\mu^2_c}=4m(m+1)\left(\frac{4+n}{3+n}\frac{1}{\Delta}-1\right).
\end{eqnarray}
Very light modes emerge if and only if $\lambda\approx\lambda_{\text{max}}(1)$
and $n$ is very large.

Finally, we shall make a brief comment on scalar perturbations in the present model.
Our $(6+n)$-dimensional description would be especially powerful
when studying scalar perturbations,
as is the case in the Randall-Sundrum-type braneworld inflation 
model with a bulk scalar field~\cite{kt}.
This is mainly
because the extra scalar degree of freedom in six dimensions
is encoded in a metric perturbation in $(6+n)$-dimensions.
This fact allows one to find a convenient gauge in which both perturbation equations
and boundary conditions become simple enough to handle.
The emergence of a tachyonic perturbation mode
in the scalar sector in the 6D Einstein-Maxwell model~\cite{ksm}
implies that such instability may arise in general $(6+n)$-dimensions,
and hence the scalar perturbations need
a detailed investigation (see also Refs.~\cite{Lee-P, Kick}
for scalar perturbations
in the Minkowski brane models in the context of Nishino-Sezgin supergravity).
This issue
will be considered carefully in a forthcoming publication.

\section{Summary}\label{final}

In this paper we have proposed a new systematic method quite useful for
studying brane cosmology in warped flux compactifications.
The crucial key is that
the Einstein-Maxwell-dilaton system in six dimensions is
equivalent to the $(6+n)$-dimensional pure Einstein-Maxwell system
under our metric ansatz.
Using this fact, we have demonstrated that
time-dependent warped compactification solutions can be obtained easily.
Our ``seed'' is essentially the double Wick rotated black hole solution
in Einstein-Maxwell theory, which we believe is more simple and familiar.
For a general dilatonic coupling $\gamma$,
we have found a power-law inflationary solution
and two nontrivial time-dependent solutions.
The former solution was turned out to be
the late-time attractor for $\lambda\neq0$.
In the case of Nishino-Sezgin supergravity ($\gamma=1$), the power-law solution
reduces to the noninflating solution of~\cite{Scaling}.
Here we have obtained the explicit form of the metric of the internal two-dimensional space.
For the particular choice of the parameter, $\lambda=0$,
we have seen that
our solution reproduces the cosmological solution of~\cite{MN1, MN2}
(but with warped bulk geometry and conical deficits,
which do not affect the geometry of the external space).

We have also investigated the dynamics of tensor perturbations using
our $(6+n)$-dimensional description.
We obtained
the separable equation of motion for axisymmetric tensor perturbations,
and showed that there always exists a zero mode,
whereas we have no tachyonic modes in the spectrum.
For relatively small $\lambda$, which is likely
from the stability consideration,
KK modes are too heavy to be excited during inflation.


\acknowledgments

We wish to thank Osamu Seto for comments.
TK is supported by the JSPS under Contract No.~19-4199.
The work of MM was supported by the project ``Transregio (Dark
Universe)" at the ASC.



\appendix
\section{Derivation of the Kasner-type metric}\label{App:Kasner}

In this appendix we present a detailed derivation of the Kanser-type metric used
in the main text.
The metric takes the form
\begin{eqnarray}
\bar g_{\alpha\beta}dz^{\alpha}dz^{\beta}=-dt^2+
\underbrace{e^{2A(t)}\delta_{ij}dx^idx^j}_{\text{3D}}
+
\underbrace{e^{2B(t)}\delta_{mn}dx^mdx^n}_{n\text{D}}.
\end{eqnarray}
The field equations $\bar R_{\alpha\beta}=[2/(2+n)]\lambda \bar g_{\alpha\beta}$ read
\begin{eqnarray}
(t\;t):\quad
3\left(\ddot A+\dot A^2\right)+n\left(\ddot B+\dot B^2\right)&=&\frac{2\lambda}{2+n},
\label{K1}
\\
(i\;j):\qquad\qquad\;\;
\ddot A+\dot A\left(3\dot A + n\dot B\right)
&=&\frac{2\lambda}{2+n},
\label{K2}
\\
(m\;n):\qquad\qquad\;
\ddot B+\dot B\left(3\dot A + n\dot B\right)
&=&\frac{2\lambda}{2+n},
\label{K3}
\end{eqnarray}
where the dot denotes a derivative with respect to $t$.
From Eqs.~(\ref{K2}) and~(\ref{K3}) we get 
$\left(e^{3A+nB}\right)^{\cdot \cdot}=\tilde\lambda e^{3A+nB}$ and
$\left[\left(\dot A-\dot B\right)e^{3A+nB}\right]^{\cdot}=0$,
where $\tilde\lambda:=[2(3+n)/(2+n)]\lambda$.
For $\lambda\neq0$, integration of these two equations gives
\begin{eqnarray}
&&e^{3A+nB}=C_1 e^{\sqrt{\tilde\lambda}t}+C_2 e^{-\sqrt{\tilde\lambda}t},
\\
&&\left(\dot A-\dot B\right)e^{3A+nB}=C_3\sqrt{\frac{(2+n)\tilde\lambda}{3n}}.
\end{eqnarray}
Substituting this result into Eq.~(\ref{K1}), we obtain
\begin{eqnarray}
4C_1C_2+C_3^2=0. \label{K_constraint}
\end{eqnarray}
If $C_3=0$, the relation~(\ref{K_constraint}) implies that $C_2=0$
(We are interested in an expanding case).
Then we have the de Sitter solution:
\begin{eqnarray}
A(t)=H_0 t +A_0, \quad B(t)=H_0t+B_0,
\end{eqnarray}
with $H_0^2=2\lambda/[(3+n)(2+n)]$.
This solution has two integration constants, $A_0$ and $B_0$.
If $C_3\neq0$, we may write $C_1=\pm(C_3/2)e^{-\sqrt{\tilde\lambda}t_0}$
and
$C_2=\mp(C_3/2)e^{\sqrt{\tilde\lambda}t_0}$, where $t_0$ is a constant. In this case we have
\begin{eqnarray}
e^{3A+nB}&=&C_3\sinh\left[
\sqrt{\tilde\lambda}(t-t_0)
\right], \label{nont1}
\\
e^{A-B}&=&C_4\left\{\tanh\left[
\frac{\sqrt{\tilde\lambda}}{2}(t-t_0)
\right]\right\}^{\pm\sqrt{(2+n)/3n}}, \label{nont2}
\end{eqnarray}
where we have introduced another integration constant $C_4$ in addition to $C_3$ and $t_0$.
This solution has three integration constants, in contrast to the de Sitter case.

In the case of $\lambda=0$ it is easy to find
\begin{eqnarray}
e^{3A+nB}&=&C_1(t-t_0),\\
e^{A-B}&=&C_2\left(t-t_0\right)^{\pm\sqrt{(2+n)/3n}}.
\end{eqnarray}
The behavior of the solution is the same as
that of~(\ref{nont1}) and~(\ref{nont2}) at $t\sim t_0$.

\section{Effective theory in the zero mode sector}\label{App:effective4d}

The effective four-dimensional action governing the induced geometry
on the brane at $\xi=1$ is
\begin{eqnarray}
S_{\text{eff}}=\frac{M_{(4)}^2}{2}\int d^4x\sqrt{-q}\left[
e^{nB/2}R^{(4)}-\omega_{\text{BD}}\frac{n^2}{4}
e^{nB/2}q^{\mu\nu}\partial_{\mu}B\partial_{\nu}B-2\lambda
\right],
\label{effective4d}
\end{eqnarray}
with $\omega_{\text{BD}}:=\frac{8+n}{2n}$.
This can be obtained
from the Einstein-Maxwell action~(\ref{6+n-d-action})
simply by substituting the ansatz
\begin{eqnarray}
\cG_{MN}d\cX^M\cX^N=\chi^2\left[\underbrace{e^{-nB(x)/2}
q_{\mu\nu}(x)dx^{\mu}dx^{\nu}}_{4\text{D}}
+\underbrace{e^{2B(x)}\delta_{mn}dy^mdy^n}_{n\text{D}}\right] + \frac{1}{2\Lambda}\left(
\frac{d\chi^2}{f}+\beta^2 fd\theta^2\right),
\end{eqnarray}
and
$\cF_{\chi\theta}=M^{2+n/2}_{(6+n)}\beta/\sqrt{2\Lambda}\cdot \hat Q/\chi^{4+n}$~[Eq.~(\ref{F-chi-th})],
with the definition of the four-dimensional Planck mass
$M_{(4)}^2:=M_{(6)}^4\int\sqrt{\cG_{\chi\chi}\cG_{\theta\theta}}d\chi d\theta$.

The conformal transformation $\tilde q_{\mu\nu}=e^{n B/2}q_{\mu\nu}$ leads to
\begin{eqnarray}
S_{\text{eff}}=\frac{M_{(4)}^2}{2}\int d^4x\sqrt{-\tilde q}\left[
\tilde R^{(4)}-\frac{n(2+n)}{2}\tilde q^{\mu\nu}\partial_{\mu}B\partial_{\nu}B-2\lambda
e^{-nB}
\right].
\label{effective4d_conf}
\end{eqnarray}
Thus the zero-mode sector of our model can be described by
the Einstein-scalar field system with an exponential potential.




\end{document}